\documentclass[aps,prl,twocolumn,amsmath,amssymb,superscriptaddress,floatfix]{revtex4-2}

\usepackage{multirow}
\usepackage{xcolor}
\usepackage{bm}
\usepackage{amsfonts,amssymb,amsmath}
\usepackage{graphicx,dcolumn,bm,color,braket,slashed}
\usepackage{times} 
\usepackage{comment}
\usepackage{array}
\usepackage{textcomp}
\usepackage{siunitx}

\newcommand{\comb}[2]{{}_{#1}\mathrm{C}_{#2}}

\renewcommand{\v}[1]{{\bf #1}}

\begin{document}

\title{Quasi-localization and Wannier Obstruction in Partially Flat Bands}

\author{Jin-Hong Park}
\affiliation{Research Center for Novel Epitaxial Quantum Architectures, Department of Physics, Seoul National University, Seoul, 08826, Korea}

\author{Jun-Won Rhim}
\email{jwrhim@ajou.ac.kr}
\affiliation{Research Center for Novel Epitaxial Quantum Architectures, Department of Physics, Seoul National University, Seoul, 08826, Korea}
\affiliation{Department of Physics, Ajou University, Suwon 16499, Korea}

\begin{abstract}
The localized nature of a flat band is understood by the existence of a compact localized eigenstate. 
However, the localization properties of a partially flat band, ubiquitous in surface modes of topological semimetals, have been unknown. 
We show that the partially flat band is characterized by a non-normalizable compact localized state(NCLS).
The partially flat band develops only in a momentum range, where normalizable Bloch wave functions can be obtained by the linear combination of the NCLSs.
Outside this momentum region, a ghost flat band, unseen from the band structure, is introduced for the consistent counting argument with the full set of NCLSs.
Then, we demonstrate that the Wannier function corresponding to the partially flat band exhibits an algebraic($\sim 1/r^{1+\epsilon}$ in 1D and $\sim 1/r^{3/2+\epsilon}$ in 2D) decay behavior, where $\epsilon$ is a positive number.
Namely, one can have the Wannier obstruction even in a topologically trivial band if it is partially flat.
Finally, we develop a construction scheme of a tight-binding model of the topological semimetal by designing an NCLS.
\end{abstract}

\maketitle

\textit{Introduction.}
Flat bands have received significant attention because they are considered an ideal playground to study many-body physics~\cite{ferro_mielke1993ferromagnetism,ferro_tasaki1998nagaoka,ferro_hase2018possibility,ferro_sharpe2019emergent,ferro_saito2021hofstadter,supercon_aoki2020theoretical,supercon_volovik1994fermi,supercon_volovik2018graphite,supercon_balents2020superconductivity,supercon_yudin2014fermi,supercon_liu2021spectroscopy,supercon_peri2021fragile,wu2007flat,chen2018ferromagnetism,jaworowski2018wigner,wang2011nearly,tang2011high,sun2011nearly,neupert2011fractional,sheng2011fractional,regnault2011fractional,liu2012fractional,bergholtz2013topological} and geometric properties of the Bloch wave function beyond topological aspects~\cite{rhim2020quantum,hwang2021geometric,hwang2021wave,hwang2021flat,supercon_peotta2015superfluidity,mera2022nontrivial,oh2022bulk,yu2023nontrivial,tian2023evidence,jung2023quantum}.
While it has long been believed that the flat band can only be realized in artificial systems~\cite{guzman2014experimental,vicencio2015observation,mukherjee2015observation,ma2020direct,milicevic2019type,taie2015coherent,drost2017topological}, there has recently been a surge of research on synthesizing flat band materials~\cite{kang2020topological,liu2020orbital,yin2019negative,kang2020dirac,lin2018flatbands,di2023flat} after discovering the unconventional superconductivity originating from the flat band in the twisted bilayer graphene~\cite{supercon_cao2018unconventional}.
As a result, flat band materials are now regarded as intriguing quantum materials with promising implications, attracting interest from both fundamental and application-oriented perspectives.

The localization of charge carriers in a flat band, essential for understanding many-body phenomena in the system, has an intriguing origin.
Suppose a flat band is perfectly dispersionless over the whole Brillouin zone, denoted by a fully flat band(FFB).
The localization nature of the FFB is understood by the existence of a special eigenmode called a compact localized state(CLS)~\cite{rhim2021singular,rhim2019classification,leykam2018artificial}.
The amplitude of the CLS is nonzero only inside a finite region while completely vanishing outside of it.
The CLS can be stabilized due to the destructive interference provided by the flat band system's specific lattice and hopping structures.
This explains how the electrons can be localized in the flat band systems, although electrons are itinerant via the hopping processes.
Moreover, it was shown that when the flat band's Bloch eigenfunction possesses a singularity due to a band-crossing, one cannot find a set of independent CLS spanning the flat band.
In this case, several extended states, called the non-contractible loop states(NLSs), should be added to the set to achieve completeness~\cite{rhim2019classification,rhim2021singular}.
NLSs exhibit exotic real-space topology probed in photonic lattices~\cite{ma2020direct}.
From designing CLSs and NLSs, one can also easily construct flat band tight-binding models~\cite{kim2023general}.

Besides the FFB, one can also observe partially flat bands(PFBs) frequently on the edge or surface of topological semimetals, such as graphene and nodal line semimetals~\cite{fujita1996peculiar,rhim2008edge,jaskolski2011edge,burkov2011topological,deng2019nodal,kim2015dirac,bzduvsek2016nodal}.
In the zigzag graphene nanoribbon(ZGNR), we have PFBs between two Dirac points.
In fact, they are nearly flat bands with momentum dependence $\sim (k-\pi)^Q$ around the zone boundary, where $Q$ is the number of dimmer lines proportional to the ribbon width.
In the semi-infinite limit($Q\rightarrow\infty$), the exact PFB is realized.
The PFBs of the ZGNR attracted tremendous attention due to their instability toward the half-metallic ground state~\cite{son2006half}.
Another well-known example is the drum-head surface states of nodal-line semimetals~\cite{burkov2011topological,deng2019nodal,kim2015dirac,bzduvsek2016nodal}, which are expected to lead to the high-temperature superconductivity~\cite{heikkila2011flat}. 

\begin{figure}[t]
\centering
\includegraphics[width=0.48\textwidth ]{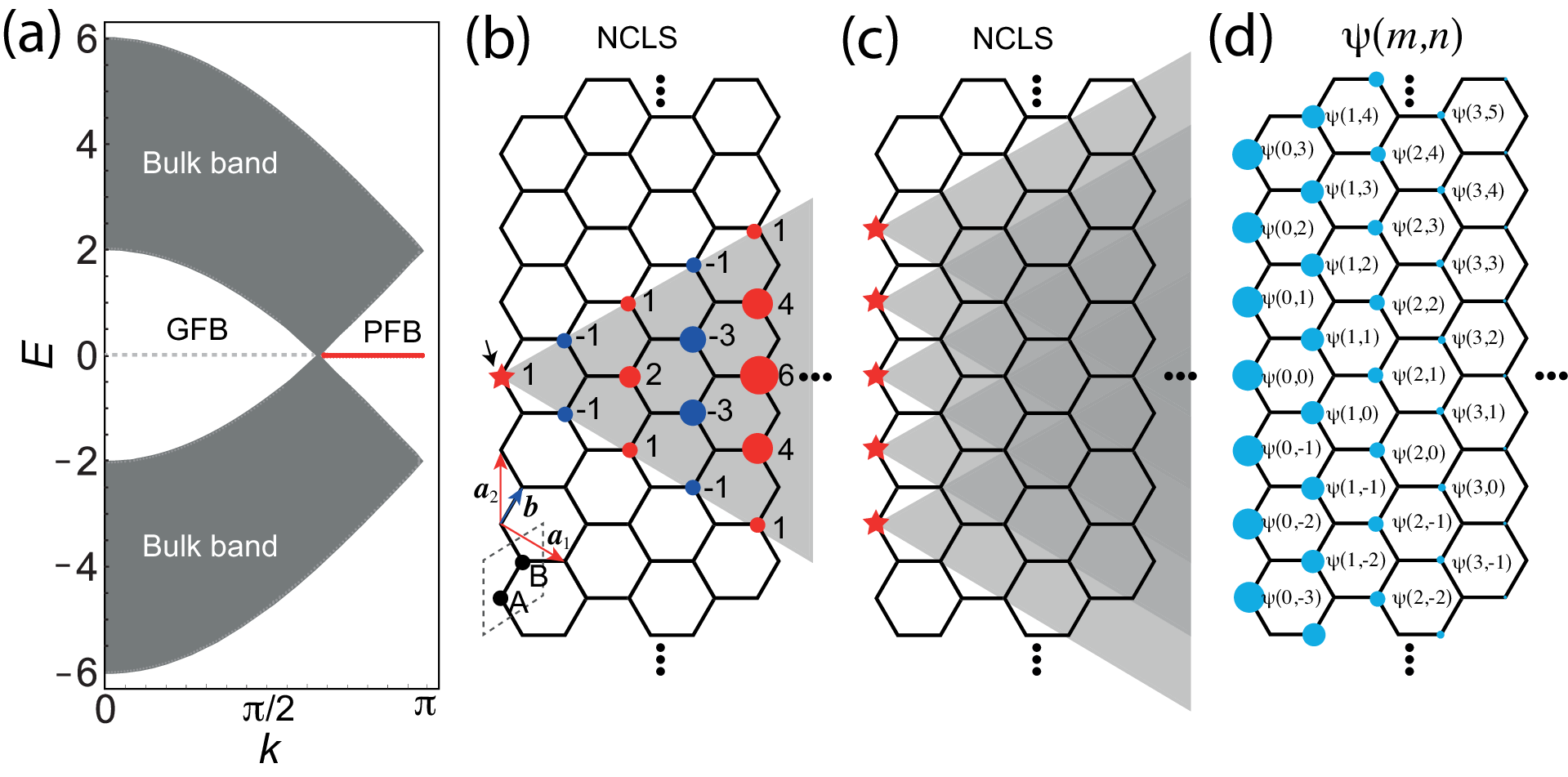}
\caption{(a) The band structure of a semi-infinite zigzag graphene nanoribbon. At zero energy, the PFB and GFB are plotted by the red solid and gray dashed lines, respectively. (b) The semi-infinite zigzag graphene nanoribbon has an edge on the left-hand side. The NCLS is drawn by a gray region. The amplitudes of the NCLS are denoted by the colored circles, whose size stands for the magnitude of the amplitude, while the color represents the sign. We have different NCLSs depending on the position of the vertex indicated by the red star symbol. (c) A linear combination of the NCLSs, resulting in a Bloch wave function, whose amplitudes are plotted in (d).
}\label{fig:grp}
\end{figure}

This letter answers the question of how to understand the localization nature of PFBs.
In the case of the FFB, one can always find $N$(the number of unit cells in the system) CLSs spanning the flat band completely if the flat band has no singularity~\cite{rhim2019classification}.
This property characterizes the localization nature of the FFB~\cite{bergman2008band,leykam2018artificial,rhim2019classification}.
The same idea may not be applied to the PFB because it has only $fN$ degenerate Bloch states, where $f$ is a fractional number($0<f<1$) representing the ratio between the flat region and the Brillouin zone. 
However, $N$ number of different CLSs must exist if at least one CLS is found, by moving the center of the CLS to different unit cells.
This macroscopic mismatch between the number of Bloch wave functions and CLSs implies that the conventional CLS cannot exist in the PFB.
%

We demonstrate that the PFB is characterized by an unconventional type of CLS called a non-normalizable compact localized state(NCLS).
%
With a normalizable regular localized eigenstate, one can find $N$ independent such states by translated copies of them.
Then, their Bloch summations with different crystal momenta give $N$ independent Bloch states, which correspond to an FFB, not a PFB.
%
The PFBs are usually found in semi-infinite systems, where the normalizability condition can be broken.
%
The NCLS shows the compactly localized feature along the translationally invariant direction.
However, its amplitude can grow away from the boundary, resulting in a non-normalizable state.
%
%
%
%
Although it does not make sense to use the non-normalizable states conventionally, we note that their Bloch sums can be normalizable depending on the crystal momentum.
%
The PFB appears in the momentum range with normalizable Bloch sums of NCLSs, while we propose that there is a ghost flat band(GFB) outside this range because the corresponding Bloch states cannot be normalized.
The unphysical, non-normalizable states cannot be observed in a spectroscopy.
However, introducing the GFB can resolve the frustrated counting argument because we have the same number of Bloch states(from both the PFB and GFB) and the NCLSs.
We also investigate the localization properties of the Wannier functions of the PFB.
The exponential decay of the Wannier function of an isolated analytic band explains the localization nature of insulators~\cite{kohn1959analytic,brouder2007exponential}.
This Wannierization is obstructed in topological bands due to the singular properties of the corresponding Bloch wave function~\cite{po2018fragile}.
We show that the Wannier obstruction can also occur in a topologically trivial band if it is a PFB because the Bloch wave functions in the GFB do not participate in the construction of the Wannier functions.
In the PFB, the Wannier function exhibits $~1/r^{1+\epsilon}$ and $~1/r^{3/2+\epsilon}$ decays with a positive number $\epsilon$ in 1D and 2D, respectively, instead of the exponential one.
Finally, We develop a systematic scheme for constructing a tight-binding Hamiltonian hosting a PFB by designing an NCLS.
Since the PFBs are usually the surface modes of topological semimetals, this scheme can be used to obtain topological semimetal models.

\textit{Non-normalizable compact localized states.}
We discuss the generic localization nature of a PFB by considering a specific example, the PFB of a semi-infinite graphene with a zigzag edge, illustrated in Fig.~\ref{fig:grp}(a).
As denoted in Fig.~\ref{fig:grp}(b), positions of A and B sites are represented by $\mathbf{R}_A(m,n) = m\mathbf{a}_1 + n\mathbf{a}_2$ and $\mathbf{R}_B(m,n) = m\mathbf{a}_1 + n\mathbf{a}_2 + \mathbf{b}$, respectively, where $\mathbf{a}_1=a(\sqrt{3}/2,-1/2)$, $\mathbf{a}_2=a(0,1)$, and $\mathbf{b}=(a/\sqrt{3})(1/2,\sqrt{3}/2)$.
Here, $m\geq 0$ is the dimmer line index, and $a$ is the lattice constant.
Here, only the nearest neighbor hopping processes are included.
We look for a CLS localized along the translationally invariant direction, namely $y$-axis in Fig.~\ref{fig:grp}(b).
%
One can find a zero energy eigenstate, illustrated in Fig.~\ref{fig:grp}(b), whose amplitudes at $A$ sites are given by
\begin{align}
    \phi^{A}_{n^*}(m.n)= \begin{cases}
        (-1)^m\comb{m}{n-n^*}, & n^*\leq n \leq n^*+m \\
        0, & \mathrm{otherwise}
    \end{cases},
\end{align}
where $n^*\mathbf{a}_2$ is the position of the left-end vertex of this state, as indicated by an arrow in Fig.~\ref{fig:grp}(b).
The amplitudes at B sites, denoted by $\phi^{B}_{n^*}(m.n)$, are zero.
In total, the eigenstate is written as $|\phi_{n^*}\rangle=\sum_{m,n}[\phi^A_{n^*}(m,n)a^\dag_{m,n}+\phi^B_{n^*}(m,n)b^\dag_{m,n}]|0\rangle$, where $a^\dag_{m,n}$($b^\dag_{m,n}$) creates an electron at A(B) site with position $\mathbf{R}_A(m,n)$($\mathbf{R}_B(m,n)$).
The compactly localized feature of this state is due to the destructive interference between $\phi^{A}_{n^*}(m,n^*)$($\phi^{A}_{n^*}(m,n^*+m)$) and $\phi^{A}_{n^*}(m+1,n^*)$($\phi^{A}_{n^*}(m+1,n^*+m)$) at the neighboring B site after the hopping processes.
However, this is a non-normalizable state since the amplitudes grow as the dimmer line index $m$ increases.
This is the NCLS described in the introduction.
There are $N$ degenerate NCLSs at zero energy depending on the vertex position $n^*$, where $N$ is the number of unit cells along the $y$-axis.
Since the left-end vertices of the $N$ NCLSs with different $n^*$'s do not overlap, the $N$ NCLSs are independent of each other even within the periodic boundary condition along $y$-axis.

\begin{figure}[t]
\centering
\includegraphics[width=0.47\textwidth ]{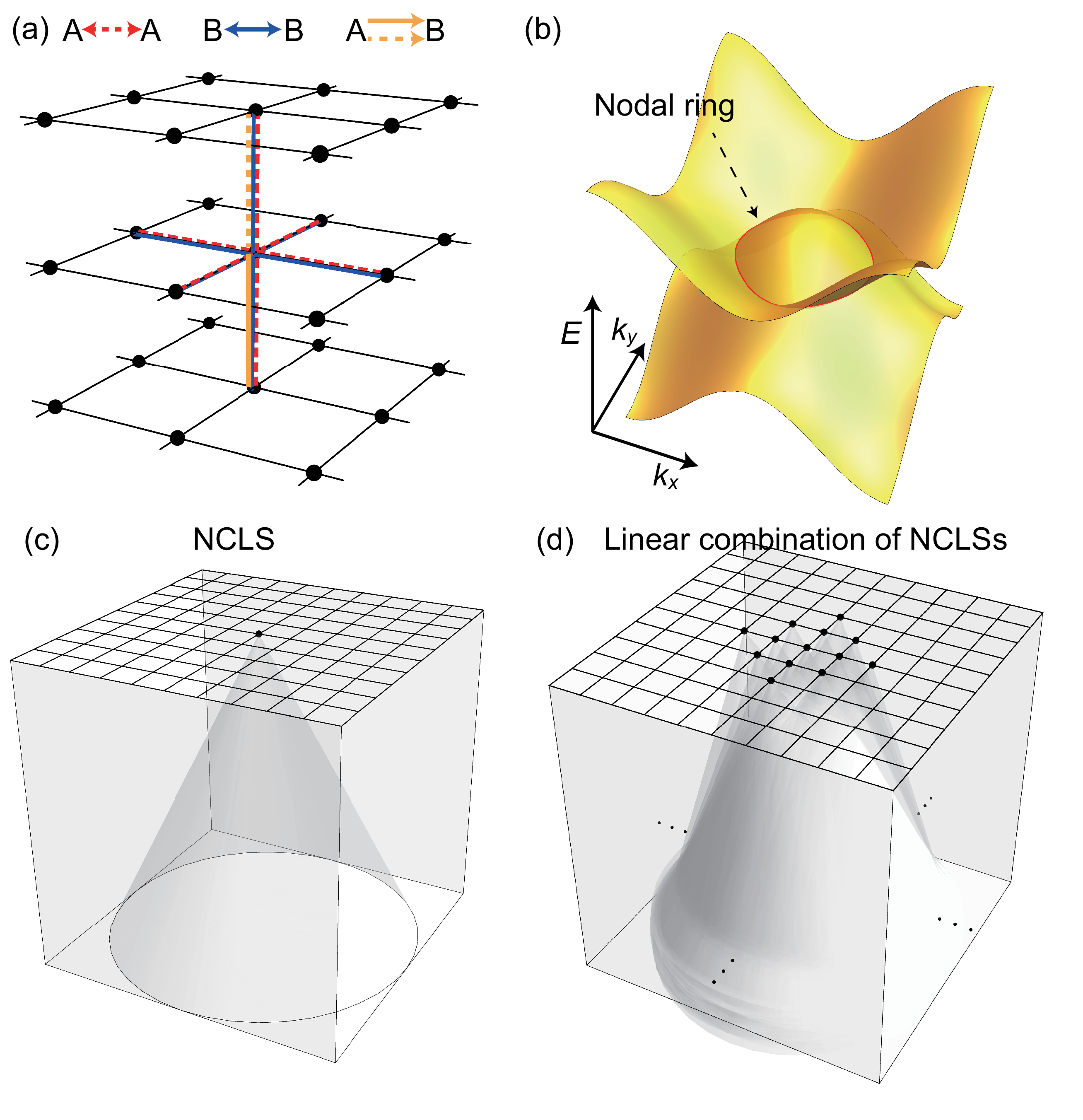}
\caption{(a) The lattice and hopping structures of the 3D nodal line semimetal model. At each site, two orbitals, denoted by A and B, reside. Lines with red, blue, and yellow colors represent hopping processes. Solid and dashed lines indicate $+$ and $-$ signs of the hopping amplitudes. (b) The band structure for $k_z=0$. The nodal ring is denoted by the red closed curve. (c) A schematic view of the NCLS of the nodal line semimetal in the semi-infinite system. (d) A linear combination of NCLSs in (c) with different apex positions, resulting in a Bloch wave function.
}\label{fig:NLS}
\end{figure}

We show that a normalizable Bloch eigenstate can be obtained from a linear combination of the NCLSs, as illustrated in Fig.~\ref{fig:grp}(c).
We build an extended eigenstate satisfying the Bloch theorem given by
\begin{align}
    |\psi_k \rangle= \sum_{n^*} e^{i n^* k} | \phi_{n^*} \rangle,
\end{align}
whose amplitude reads
\begin{align}
    \psi^A_k(m,n) = e^{ink} (1+e^{-ik})^m, \label{eq:grp+bloch}
\end{align}
for the A sites at $\mathbf{R}_A(m,n)$ while $\psi^B_k(m,n)=0$, as plotted in Fig.~\ref{fig:grp}(d).
The eigenenergy corresponding to this mode is zero because that of $| \phi_{n^*} \rangle$ is zero.
Most importantly, the obtained Bloch eigenstate shows exponential decay along $\mathbf{a}_2$ direction if $k \in I_L=(-\pi,-2\pi/3)$ or $k \in I_R=(2\pi/3,\pi)$.
Here, the Brillouin zone is set to be $-\pi<k\leq \pi$.
Namely, $|\psi_k \rangle$ is a normalizable one-dimensional Bloch wave function for momenta belonging to $I_L$ or $I_R$.
Outside these two intervals, on the other hand, the constructed Bloch states should be discarded and invisible in spectroscopy because they cannot be normalized.
Therefore, we have zero energy flat band in $I_L$ and $I_R$, which realizes the PFB.
This PFB is not connected to other bands and is terminated in the middle of the Brillouin zone.
Meanwhile, one can consider the zero energy states corresponding to the non-normalizable $|\psi_k \rangle$ outside $I_L$ and $I_R$ as a ghost flat band(GFB).
We have in total $N$ number of states in the PFB and GFB, which is consistent with the $N$ number of independent NCLSs $| \phi_{n^*} \rangle$.

In general, we conjecture that a PFB is obtained in a semi-infinite system by discarding the ghost part of an FFB.
%
%
A semi-infinite system is a good playground to have such a ghost band for the following reasons.
Considering only hermitian tight-binding Hamiltonians, which can be block-diagonalizable into the finite-size Bloch Hamiltonians, the bulk's band dispersion should be analytic in momentum because every matrix element of the Bloch Hamiltonian of such a model is analytic in momentum space.
However, the PFB cannot be an analytic function of momentum because an integer $n$ exists such that the $n$-th left- and right-hand derivative of the band becomes discontinuous at the end-point of the PFB.
Therefore, the flat band should always be FFB in such a conventional Hamiltonian.
One of the ways to violate the properties of the conventional Hamiltonians is to make the size of the Bloch Hamiltonian infinite, which is impossible in reality in principle.
The semi-infinite systems are the examples, and the non-normalizable states can emerge in such systems, as shown in the example above.

We apply our general conclusions to a 3D topological nodal-line semimetal.
We consider the minimal tight-binding Hamiltonian on a cubic lattice given by $H_\mathrm{NL}=\sin k_z \sigma_y + (2-\cos k_x - \cos k_y - \cos k_z)\sigma_z$, where $\sigma_\alpha$ is a Pauli matrix representing the orbital degree of freedom for each site.
The hopping structure is illustrated in Fig.~\ref{fig:NLS}(a).
Lattice sites are represented by $\mathbf{R}(\mathbf{n})=n_x \mathbf{a}_x + n_y \mathbf{a}_y +n_z \mathbf{a}_z$, where $\mathbf{a}_\alpha$ is a unit vector along $\alpha$-direction.
This model hosts a nodal ring described by $1=\cos k_x + \cos k_y$ at zero energy, as plotted in Fig.~\ref{fig:NLS}(b).
When the system is terminated along $z$-direction, we have a drumhead flat surface mode inside the nodal ring.
We consider a semi-infinite system obtained by removing sites with a negative $n_z$ in the cubic lattice.
The quasi-localized feature of this drumhead PFB is explained by a zero energy NCLS illustrated in Fig.~\ref{fig:NLS}(c), whose apex is located at $\mathbf{R}(\mathbf{n}^*)=n^*_x\mathbf{a}_x + n^*_y\mathbf{a}_y$.
The amplitude of the NCLS, denoted by $|\phi_{\mathbf{n}^*}\rangle$, is given by
\begin{align}
    \phi_{\mathbf{n}^*,\sigma}(\mathbf{n}) = \sigma \Big[ 2-\sum_{\beta}\left(e^{ik_\beta}+e^{-ik_\beta}\right)/2\Big]^{n_z}\Big|_{\Delta n_x,\Delta n_y},
\end{align}
where $\sigma=\pm$ represents two orbitals, $\beta=x$ and $y$, and the subscript $\Delta n_\beta = n_\beta - n_\beta^*$ means that we pick up the coefficient of the exponential factor $e^{i\Delta n_x k_x + i\Delta n_y k_y}$.
Since the above formula is a finite polynomial of the exponential factors, $\phi_{\mathbf{n}^*,\sigma}(\mathbf{n})$ is nonzero only inside a finite region for a given $n_z$, which implies the compact localization.
Then, by a linear combination of the NCLSs(Fig.~\ref{fig:NLS}(d)), we obtain Bloch eigenfunction, $|\psi_\mathbf{k} \rangle= \sum_{\mathbf{n}^*} e^{i\mathbf{n}^*\cdot\mathbf{k}} |\phi_{\mathbf{n}^*}\rangle$, whose amplitudes are evaluated as
\begin{align}
    \psi_{\mathbf{k},\sigma}(\mathbf{n}) = e^{i(n_xk_x+n_yk_y)} (2-\cos k_x - \cos k_y)^{n_z},
\end{align}
where $\mathbf{k}=(k_x,k_y)$.
The Bloch wave function exhibits exponential decay and can be normalized in a momentum range $-1< 2-\cos k_x -\cos k_y <1$, precisely the same as the drum-head surface mode's region above.
Outside the nodal ring, the constructed Bloch wave function increases exponentially as a function of $n_z$ and cannot be normalized, which leads to the GFB.

\textit{Wannier obstruction.}
The localization nature of an insulator is explained by exponentially decaying Wannier functions~\cite{brouder2007exponential}. 
According to the Paley-Wiener theorem, such states are guaranteed to exist if the Bloch wavefunctions are analytic over the whole Brillouin zone~\cite{des1964analytical}.
Bloch wavefunctions are always analytic in 1D, and one can find exponentially localized Wannier functions in a 1D isolated band.
On the other hand, in 2D, a topologically nontrivial band cannot have such a Wannierization due to singularities of the corresponding Bloch eigenfunction in momentum space.
Here, we investigate the Wannier function properties of a PFB, where one cannot apply the Paley-Wiener theorem because the Fourier transform of the Bloch wave function is ill-defined due to the absence of the wave function outside the PFB regions in the Brillouin zone.

Let us begin with the 1D PFB of a semi-infinite system.
A general form of the Bloch wave function of the PFB is given by
\begin{align}
    |\psi_{k}\rangle = \lim_{Q\rightarrow\infty}\frac{1}{\sqrt{N}}\sum_{m=1}^Q \sum_{\sigma_m=1}^{P_n} v^{(Q)}_{m,\sigma_m}(k) e^{ikn} c^\dag_{m,n,\sigma_m}|0\rangle,
\end{align}
where $n$ is the unit cell index along the translationally invariant direction, $m$ is the sublattice index, and $\sigma_m$ represents the orbital in the $m$-th sublattice.
Here, $m$ increases as we go away from the semi-infinite system's boundary($m=0$).
The semi-infinity is described by the limit process $Q\rightarrow \infty$.
We assume that there are $P_m$ orbitals in the $m$-th sublattice.
Finally, $v^{(Q)}_{m,\sigma_m}(k)$ is the component corresponding to the $\sigma_m$-th orbital in the $m$-th sublattice of the PFB's eigenvector $\v v^{(Q)}(k)$ of the Bloch Hamiltonian.
Here, $\v v^{(Q)}(k)$ is obtained by holding the components of the unnormalized Bloch wave function constructed from the NCLSs, such as (\ref{eq:grp+bloch}), up to the $Q$-th sublattice and normalizing it.
For example, for the semi-infinite graphene, we have $v^{(Q)}_{m}(k) = \psi^A_k(m,0)/(\sum_{m^\prime=0}^Q |\psi^A_k(m^\prime,0)|^2)^{1/2}$.
Note that, before reflecting the semi-infinity($Q\rightarrow \infty$) of the system, $v^{(Q)}_{m,\sigma_m}(k)$ is analytic over the whole Brillouin zone, even out of the PFB region. 
As a result, the Fourier transform of $v^{(Q)}_{m,\sigma_m}(k)$ exponentially decays far from the Wannier center according to the Paley-Wienener theorem and we have $v^{(Q)}_{m,\sigma_m}(k) = (1/\sqrt{N})\sum_n e^{ikn} \varphi^{(Q)}_{m,\sigma_m}(n)$, where $\varphi^{(Q)}_{m,\sigma_m}(n) \sim e^{-b|n|}$ for $|n|\rightarrow\infty$ with a positive coefficient $b$.
Now, we obtain the Wannier function for the PFB by using this form of $v^{(Q)}_{m,\sigma_m}(k)$ in the semi-infinite limit($Q\rightarrow \infty$).
In this limit, $v^{(Q)}_{m,\sigma_m}(k)$ is ill-defined for momenta corresponding to the EFB.
Therefore, performing the integral over the momentum range corresponding to the PFB is natural to construct a Wannier function from $v^{(Q)}_{m,\sigma_m}(k)$.
Then, the Wannier function amplitude for the $\sigma_m$-th orbital is given by $W_{m,\sigma_m}(n-n_0) = (1/N)\lim_{Q\rightarrow\infty}\sum_{k\in I_\mathrm{PFB}} v^{(Q)}_{m,\sigma_m}(k)e^{ik(n-n_0)}$, where $I_\mathrm{PFB}$ is a set of momenta supporting the PFB, and $n_0$ is the Wannier function's center.
In general, $I_\mathrm{PFB}$ is divided into many separate intervals, where the $l$-th one is $[k_{l,i},k_{l,f}]$.
%
Denoting $\Delta n=n-n_0$, $v_{m,\sigma_m}(k)=\lim_{Q\rightarrow\infty}v^{(Q)}_{m,\sigma_m}(k)$, and $\varphi_{m,\sigma_m}(k)=\lim_{Q\rightarrow\infty}\varphi^{(Q)}_{m,\sigma_m}(k)$, the Wannier function of the PFB is given by
\begin{align}
    W_{m,\sigma_m}(\Delta n) = \frac{1}{2\pi i}\sum_{l,\alpha} s_\alpha\frac{e^{ik_l\Delta n}}{\Delta n}F_{l,m,\sigma_m}(\Delta n),\label{eq:wannier}
\end{align}
where
\begin{align}
    F_{l,m,\sigma_m}(\Delta n) = \frac{1}{\sqrt{N}}\sum_{n^\prime} \varphi_{m,\sigma_m}(n^\prime)\frac{e^{ik_{l,\alpha}}}{n^\prime/\Delta n +1 },\label{eq:wannier_factor}
\end{align}
and $s_\alpha=-1$ or $1$ for $\alpha=i$ or $f$, respectively.
Since we are interested in the behavior of $W(\Delta n)$ in the limit $|\Delta n|\rightarrow\infty$, let us consider a function $\tilde{F}_{l,m,\sigma_m}(x) = F_{l,m,\sigma_m}(1/x)$, where $x=1/\Delta n$.
One can show that $\tilde{F}_{l,m,\sigma_m}(0) = 0$ because the eigenvector component $v_{m,\sigma_m}(k)$ should be zero at $k_{l,i}$ and $k_{l,f}$, boundaries of the PFB to be continuously connected to the eigenvector component of the EFB, which is zero in the limit $Q\rightarrow\infty$.
To figure out the functional form of $\tilde{F}_{l,m,\sigma_m}(x)$ near $\alpha=0$, we consider an arbitrary order of derivatives of it at $\alpha=0$.
We evaluate $\partial_x^p \tilde{F}_{l,m,\sigma_m}(x)|_{x=0}$ and denote the smallest $p$, where the derivative becomes nonzero, by $p^*$.
If $|\partial_x^{p} \tilde{F}_{l,m,\sigma_m}(x)|_{x=0}|<\infty$ for $p\geq p^*$, one can deduce that $\tilde{F}_{l,m,\sigma_m}(x) \sim x^{p^*}$.
This implies that $W_{m,\sigma_m} \sim 1/r^{1+p^*}$ from (\ref{eq:wannier}).
On the other hand, $\partial_x^{p^*} \tilde{F}_{l,m,\sigma_m}(x)|_{x=0}$ can be divergent too.
In this case, one can infer that $\tilde{F}_{l,m,\sigma_m}(x)$ is an irrational function of the form $x^{\epsilon}$, where $p^*-1<\epsilon <p^*$ because $\tilde{F}_{l,m,\sigma_m}(x)$ vanishes at $x=0$.
In this case, we have $W_{m,\sigma_m} \sim 1/r^{1+\epsilon}$.
In total, the Wannier function decays algebraically as $\sim 1/r^{1+\epsilon}$, where $p^*-1 < \nu \leq p^*$.
%
%
Note that the Wannier functions with different $n_0$ of the PFB are not orthogonal to each other in general because not all the Bloch wave functions in the Brillouin zone participate in constructing the Wannier functions.
As a result, while there can be $N$ different Wannier functions with distinct $n_0$ values, only $N \sum_l(k_{l,f}-k_{l,i})/2\pi$ number of them are independent of each other.
Similar conclusions to the 1D Wannierization above can be applied to the two-dimensional PFBs.
Specifically, we show that the Wannier function corresponding to the 2D PFB exhibits $1/r^{3/2+\epsilon}$ decay far from its center, where $\epsilon$ is a positive number.
Detailed derivations are included in Supplementary information.

\begin{figure}[t]
\centering
\includegraphics[width=0.5\textwidth ]{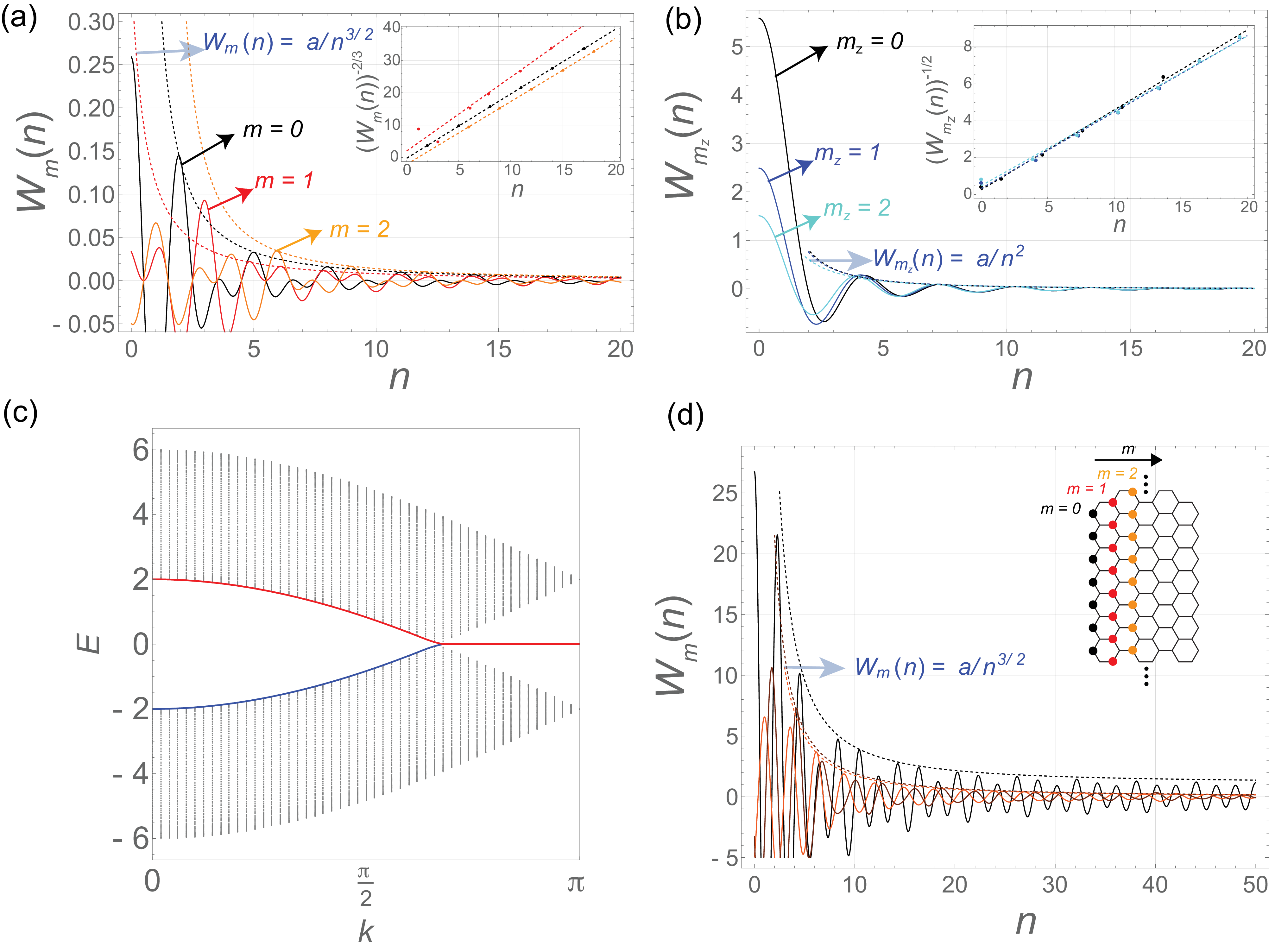}
\caption{(a) and (b) plot the Wannier function amplitudes of the semi-infinite zigzag graphene nanoribbon and nodal line semimetal, respectively. $m$ and $m_z$ represent the dimer lines and layers away from the edge and surface of these two systems. $n$ is an index of the unit cell along a translationally invariant direction. In the insets, we show that the local maxima of the Wannier functions decay algebraically. (c) A band dispersion of a finite-width zigzag graphene nanoribbon. The red and blue bands in the middle correspond to the edge states. (d) The Wannier function for the red band in (c) for various values of $m$.
}\label{fig:wan}
\end{figure}

Let us consider semi-infinite graphene as an example.
Using (\ref{eq:grp+bloch}), we have the normalized eigenvector component of the PFB, evaluated as $v_{m}(k) = \lim_{Q\rightarrow\infty} v^{(Q)}_{m}(k) = [1-4\cos^2(k/2)]^{1/2}(1+e^{-ik})^m$, where the orbital index $\sigma_m$ is dropped because we have only one orbital per sublattice.
The momentum range for the PFB is given by $I_\mathrm{PFB}=[2\pi/3,4\pi/3]$.
Outside the PFB region, $v_{m}(k) =0$.
The Wannier function is obtained by $W_m(n-n_0)=(1/2\pi)\int_{2\pi/3}^{4\pi/3}dk~ v_m(k) e^{ik(n-n_0)}$ for the $m$-th dimer line.
When $n_0=0$, the Wannier function amplitudes for several values of $m$ are plotted in Fig.~\ref{fig:wan}(a).
We note that $W_m(n)\sim 1/n^{3/2}$ for large $n$ as shown in the inset of Fig.~\ref{fig:wan}(a).
This is consistent with that $\partial_k v_m(k)$ diverges at $k=2\pi/3$ and $4\pi/3$.
This implies that the Wannier function of the semi-infinite graphene with a zigzag edge decays algebraically($\sim 1/r^{3/2}$) along the translationally invariant direction.
In the case of the semi-infinite nodal line semimetal described by $H_\mathrm{NL}$ above, the normalized eigenvector's component of the PFB corresponding to the $n_z$-th layer from the surface is given by $v_{m_z}(\mathbf{k}) = \lim_{Q\rightarrow\infty} v_{m_z}^{(Q)}(\mathbf{k}) = [(\cos k_x + \cos k_y -1)(3-\cos k_x - \cos k_y)]^{1/2}(2-\cos k_x - \cos k_y)^{n_z}$ inside the nodal ring.
Outside the nodal ring, $\lim_{Q\rightarrow\infty} v_{m_z}^{(Q)}(\mathbf{k}) =0$.
By using it, we obtain the Wannier function for the PFB, and observe another algebraic decay proportional to $1/r^2$ along various directions on the surface, as plotted in Fig.~\ref{fig:wan}(b).
This is consistent with the generic $1/r^{3/2+\epsilon}$ decay of the 2D PFB Wannier function shown above.

\textit{Construction of topological semimetal models.}
By reversely using the fact that the eigenvector's components of the PFB are in the form of the polynomial of exponential factors, one can find a topological semimetal model.
Let us describe the construction scheme by considering an example, a 2D semi-infinite rectangular lattice consisting of two sublattices denoted by $A$ and $B$ and translationally invariant along $x$ direction.
We arrange the PFB's eigenvector component as $\mathbf{v}(k) = (v_{1A}(k),v_{1B}(k),\cdots,v_{mA}(k),v_{mB}(k),\cdots)^\mathrm{T}$, where $m$ is the dimerline index from the edge.
We consider $v_{mA}(k)=(-1-a e^{ik}-a e^{-ik})^{m-1}$ and $v_{mB}(k)=0$ as a zero energy PFB's eigenvector.
One can note that this choice of the eigenvector corresponds to a NCLS, which has nonzero amplitudes only in a finite region for each given $m$.
An example of the Hamiltonian matrix of the semi-infinite system satisfying $H_{\mathrm{SI}}\mathbf{v}(k)=0$ is given by
\begin{align}
    H_\mathrm{SI}=\begin{pmatrix}
    0 & h^*_k & 0 & 0 & \cdots \\
    h_k & 0 & 1 & 0 & \cdots \\
    0 & 1 & 0 & h^*_k & \cdots \\ 
    0 & 0 & h_k & 0 & \cdots \\
    \vdots & \vdots & \vdots & \vdots & \ddots 
    \end{pmatrix},
\end{align}
where $h_\mathbf{k} = 1+a e^{ik}+a e^{-ik}$.
$\mathbf{v}(k)$ is normalizable for $\pi/2 < k < 3\pi/2$ when $a<1$ and $\pi/2 < k < \cos^{-1}(-1/a)$ or $ 2\pi-\cos^{-1}(-1/a) < k < 3\pi/2$ when $a \geq 1$.
Since the boundaries of this momentum interval correspond to the band-crossing points, the corresponding bulk Hamiltonian $H_\mathrm{bulk} = (1+2a\cos k_x + \cos k_y)\sigma_x + \sin k_y \sigma_y$, where $\sigma_\alpha$ is the Pauli matrix, leads to a topological semimetal with four Dirac nodes.

\textit{Discussion.}
By considering semi-infinite systems, we have shown that the PFB appears in a momentum range, where the corresponding Bloch summations of the NCLSs are normalizable.
As a result, when we construct Wannier functions, the integration range of the Bloch wave functions in the momentum space is smaller than the entire Brillouin zone, leading to the algebraically decaying behavior. 
Therefore, we have demonstrated that the Wannier obstruction can happen even in a topologically trivial band.

While we have only considered semi-infinite systems, one might wonder, regarding realistic systems, whether the Wannier obstruction can be applied to finite systems or not.
Here, the term finite system implies that the system's size is finite along the directions with broken translational symmetry.
Graphene nanoribbons with finite width are examples.
In such finite systems, where one can construct regular finite-size Bloch Hamiltonians, the GFB cannot exist, and each edge band should be smoothly connected to one of the bulk bands.
Therefore, the conventional Wannierization with exponentially localized wave packets should be achieved in these systems.
However, we show that the algebraic decaying behavior dominates the Wannier function near its center as the system size increases.
This is because the edge modes are unaffected by the system size, while the bulk mode's amplitudes diminish as the size grows.
As a result, the Wannier function decays exponentially only extremely far from its center, where the bulk wave function's amplitude becomes comparable to that of the Wannier function.
As an example, we consider a finite-size zigzag graphene nanoribbon.
Its band structure is drawn in Fig.~\ref{fig:wan}(c) for the ribbon width $Q=$, where the red and blue bands host nearly flat bands, corresponding to the edge modes, near the zone boundary.
By applying a small perturbation, the red and blue bands are detached so that one can construct the Wannier function for them properly.
Then, the red band's Wannier function is plotted in Fig.~\ref{fig:wan}(d) for various dimer lines, which exhibit the expected algebraic decay.

\end{document}